\documentclass{JHEP3}

\usepackage{epsfig}
\usepackage{amsmath}
\usepackage{amssymb,amsfonts}
\usepackage{graphicx}
\usepackage{graphics}
\usepackage{amsthm}
\usepackage{graphicx}
\usepackage{multirow}
\newcommand{\cA}{{\cal A}}
\newcommand{\cAb}{{\overline{\cal A}}}
\newcommand{\cF}{{\cal F}}
\newcommand{\cFb}{{\overline{\cal F}}}
\newcommand{\cD}{{\cal D}}
\newcommand{\cDb}{{\overline{\cal D}}}
\newcommand{\cQ}{{\cal Q}}

\newcommand{\cU}{{\cal U}}
\newcommand{\cN}{{\cal N}}
\newcommand{\cUb}{{\overline{\cal U}}}
\newcommand{\etab}{{\overline{\eta}}}
\newcommand{\psib}{{\overline{\psi}}}
\newcommand{\Tr}{{\rm Tr\;}}

\newcommand{\vn}{ {\bf n} }
\newcommand{\KD}{{K\"{a}hler--Dirac }}
\newcommand{\hatbmu}{\widehat{\boldsymbol {\mu}}}
\newcommand{\hatbe}{\widehat{\boldsymbol {e}}}

\def\nn{\nonumber}
\def\bec{\begin{center}}
\def\eec{\end{center}}
\def\beq{\begin{equation}}
\def\eeq{\end{equation}}
\def\bea{\begin{eqnarray}}
\def\eea{\end{eqnarray}}

\title{Phase Structure of Lattice $\cN=4$ Super Yang-Mills}

\author{Simon Catterall \\
Department of Physics, Syracuse University, Syracuse, NY 13244, USA
}

\author{Poul H. Damgaard \\
Niels Bohr International Academy and Discovery Center,
Niels Bohr Instiutute, Blegdamsvej 17, DK-2100 Copenhagen, Denmark}

\author{Thomas DeGrand\\
Department of Physics, University of Colorado, Boulder, CO 80309 USA}

\author{Richard Galvez \\
Department of Physics, Syracuse University, Syracuse, NY 13244, USA
}

\author{Dhagash Mehta \\
Department of Physics, Syracuse University, Syracuse, NY 13244, USA
}

\abstract{We make a first study of the phase diagram of four-dimensional $\cN=4$ super
Yang-Mills theory regulated on a space-time lattice.
The lattice formulation we employ is both gauge invariant and retains at all lattice spacings
one exactly preserved supersymmetry charge. Our numerical results are consistent with the
existence of a single deconfined phase at all observed values of the bare coupling.
}


\begin{document}

\section{Introduction}

Supersymmetric Yang-Mills theories are interesting from a variety
of perspectives; as four-dimensional toy models for understanding theories
such as QCD, as potential theories of Beyond Standard Model physics, and because of the
AdS/CFT correspondence, they may reveal connections to quantum gravity and string theory.
Many features of these theories, such as, for example, dynamical supersymmetry breaking,
are inherently non-perturbative in nature. This serves as strong motivation to
attempt to study such theories numerically (and perhaps also analytically) when
regularized by a space-time lattice.

Unfortunately, it has historically proven almost prohibitively difficult to
discretize supersymmetric gauge theories using traditional methods. This stems from the
fact that the supersymmetry algebra is an extension of the usual Poincar\'e algebra
and hence is broken completely by na\"ive discretization on a space-time lattice.
Even a discretized subalgebra of the full supersymmetry algebra where translations
are restricted to discrete steps cannot be
retained in general on discretized space-time. This is directly tied to
the failure of the Leibniz rule to hold for finite difference operators.
However, recently the development of a series of new theoretical tools have enabled
us to construct certain supersymmetric theories on the lattice while retaining
absolute preservation of a very small {\em subset} of the continuum
supersymmetries - see the
reviews \cite{Kaplan:2003uh, Giedt:2006pd, Catterall:2009it, arXiv:1110.5983} and
references therein. Other recent complementary approaches to the problem of exact
lattice supersymmetry can be found in \cite{Sugino:2003yb, Sugino:2004qd,
hep-lat/0507029, arXiv:0707.3533, Kanamori:2008bk, Hanada:2009hq, Hanada:2010kt,
Hanada:2010gs, Hanada:2011qx} while complementary approaches to
${\cal N}=4$ Yang-Mills using large N reduction techniques have been studied in
\cite{Ishiki:2009sg,Ishii:2008ib,Ishiki:2008te,Nishimura:2009xm,Honda:2010nx}. Retaining only a small subset of all continuum
supersymmetries may not have as many limitations as one might naively have
guessed, since the combination of exact gauge invariance and some exactly
preserved supersymmetries turn out to put severe constraints on the lattice
theories. Perhaps a useful analog is the breaking of full Euclidean rotational
invariance on a hypercubic lattice: in the continuum limit full Lorentz
symmetry is recovered, even though only a very small discrete subgroup
is preserved at any finite lattice spacing.

One way to understand the new constructions is to realize that they correspond to
discretizations of topologically twisted forms of the target continuum theories.
The associated lattice theories are possible only
if the continuum supersymmetric Yang-Mills
theories possess sufficient extended supersymmetry; the precise
requirement is that the number of supercharges must be an integer multiple of $2^D$ where
$D$ is the space-time dimension. In four dimensions this singles out a unique theory
which can be studied using these lattice constructions: $\cN=4$ super Yang-Mills theory,
the focus of the current work.

We do not wish to hide the fact that the lattice theories with exactly preserved
supersymmetries have features that, superficially, seem unusual to lattice
practitioners. The fermionic and bosonic degrees of freedom are distributed
in a somewhat unorthodox manner on the lattice. But once this thin layer of
disguise has been peeled off, {\em we are here dealing with a quite standard
lattice gauge theory, amenable to the usual tools of the field}. In many
ways, this lattice theory is quite simple.
In fact, as compared to essentially all other lattice
formulations with fermions, absence of fermionic doublers is trivial.
More precisely: the lattice beautifully arranges itself so that the
doublers are {\em needed} and essential for the exactly preserved
supersymmetry. The number of degrees of freedom, including the doublers, match
perfectly. The fermion action that arises in these constructions can
be interpreted as a Dirac-K\"{a}hler action which has long been known to be
equivalent, at the level of free field theory, to a (reduced) staggered fermion
action. It is the twisting process that
leads to the emergence of Dirac-K\"{a}hler fermions. This lattice construction is also
interesting from a mathematical point of view since it corresponds to a finite system with an
exact Q-cohomology which in turn may allow for rigorous results to be derived for
supersymmetric gauge theories.

At the intuitive level we can understand these lattice constructions from the
fact that they are invariant under only {\em one} of the supersymmetry charges.
Because of its Grassmann nature, it squares to zero. Indeed, this charge can
be viewed as a nilpotent BRST-charge associated with arbitrary field deformations,
just as in the continuum topological theory\footnote{In special cases, the
supersymmetric lattice can, correspondingly, also be invariant under another
nilpotent charge, the anti-BRST generator. This is not the case for the ${\cal N}=4$
theory.}. The anticommutator of two supersymmetry charges is by virtue of the supersymmetry
algebra a generator of space-time translations, and this is the basic obstacle
of lattice-regularized supersymmetry, where only discrete translations are
retained. By preserving only one supersymmetry charge on the lattice this
problem is removed. However what still needs to be demonstrated is that the
conservation of just one exact lattice supersymmetry is sufficient to recover
full supersymmetry in the continuum limit. Although only global symmetries
are at stake here, this is not as trivial as it may sound since counterterms
permitted by the single preserved charge may violate invariance under the
remaining supersymmetries one would hope to recover in the continuum.
There are also numerical issues: not all lattice-discretized theories are
amenable, at present, to numerical simulations due to sign problems. Indeed,
the theory we wish to study here potentially has a sign problem, and  it
is crucial to investigate its severity.

There are many other interesting issues. For example, the theory we study
here has flat directions associated with the scalar fields at the classical level. Do these flat
directions cause numerical instabilities in the actual simulations? And if so,
do they indicate that the functional integral even of the continuum theory
may be ill-defined?

 For the first time we are now in a position where we
can take seriously the path integral of supersymmetric field theories in the
sense that we can consider its discretized version in a setting that is
entirely well-defined. It is a theory defined in a finite four-volume $V$ and
with an ultraviolet cut-off given by the lattice spacing $a$. In a certain
sense, we evaluate observables in this supersymmetric theory by explicit
integration in the functional integral rather than relying on formal
manipulations that ignore issues such as convergence.

Motivation for studying this particular supersymmetric theory comes,
of course, also from the fact that it is a truly remarkable four-dimensional
gauge theory; it has highly non-trivial interactions, but it
is scale-free and it retains (super)conformal invariance even at the quantum
level. With this new lattice formulation we can explicitly study this theory
for all values of the
bare lattice coupling. In the formal continuum theory, there are well-known
analytical predictions for, $e.g.$, Wilson loops at strong coupling based on
the AdS/CFT correspondence \cite{Maldacena:1998im}. It is clearly of interest
to compare these predictions with a lattice-regularized analog of this
theory which can be studied numerically. A striking fact that we will
have to face in the simulations is that the lattice theory may also be conformal
at all couplings. This is a unique situation, which has not been
encountered before in the history of this field. Indeed, intuition might
suggest that the lattice theory could develop a strongly coupled phase
at some finite bare coupling, a phase that would have no continuum
limit, would have a mass gap, would perhaps be confining and could
undergo spontaneous chiral symmetry breaking.

In this first exploratory study of the lattice-regularized theory we
wish to probe these fundamental questions: What is the lattice phase diagram
of this theory? Is it really scale-free and conformal at all bare couplings?
Does the phase of the Pfaffian resulting from integration over the
fermionic degrees of freedom lead to a sign problem? Do the flat
directions for the scalars cause numerical instabilities? On the lattice
we will also need to regularize both the scalar flat directions (a soft breaking
of supersymmetry by means of a mass term) and the fermions (they have zero modes
that can be removed by super-symmetry breaking anti-periodic boundary
conditions or, alternatively, by fermionic mass terms).
To what extent is supersymmetry nevertheless conserved
in our actual numerical simulations? As we shall see, we find encouraging
results in all directions. Our results indicate that the $U(2)$ theory has no genuine
sign problem and that, therefore, phase quenched simulations are
justified. We see no evidence of instabilities associated with the flat directions.
The bosonic
regulator mass can be tuned to zero, and simple supersymmetric Ward
Identities are obeyed to high numerical accuracy. Moreover, and most
surprisingly, we find
no evidence of a strongly coupled phase in this lattice theory. It appears
that the perfect pairing between bosonic and fermionic degrees of
freedom imposed by just one supersymmetry generator is sufficient
to keep the theory in a scale-free deconfined phase throughout,
as in the continuum.

A brief outline of our paper is as follows. In the next section, we briefly
review the topological twisting of $\cN = 4$ super Yang-Mills theory
in the continuum. We discuss the analogous lattice procedure in
Section 3, with emphasis on how easily this twisted theory fits into
the well-known and existing framework of lattice gauge theory. We present our
numerical results in Section 4. Here, because our lattices are still
relatively small, we focus on local observables, all of which are well known in
the lattice context for studying the phase diagram of a gauge theory. We end
with some concluding remarks in Section 5.

\section{Twisted Supersymmetric $\cN=4$ Yang-Mills Theory}

As discussed in the Introduction, it is possible to discretize a class of
continuum supersymmetric Yang-Mills
theories using ideas based on topological twisting\footnote{Note
that the lattice actions constructed using the orbifolding and twisted methods
are equivalent \cite{Unsal:2006qp, Damgaard:2007xi, Catterall:2007kn,Takimi:2007nn}. Indeed the
original orbifold construction of this theory constitutes an independent UV complete construction of the
Marcus/GL twist of ${\cal N}=4$ Yang-Mills}.
Though the basic idea of twisting goes back to Witten in his seminal paper on
topological field theory \cite{Witten:1988ze}, it had actually been
anticipated in earlier work on staggered fermions on the lattice
\cite{Elitzur:1982vh}. In our context, the idea of twisting is to decompose
the fields of a Euclidean supersymmetric Yang-Mills
theory in $D$ space-time dimensions in
representations not of the original (Euclidean) rotational symmetry $SO_{\rm rot}(D)$,
but a twisted rotational symmetry, which is the diagonal subgroup of this symmetry
and an $SO_{\rm R}(D)$ subgroup of the R-symmetry of the theory, that is,
\beq
SO(D)^\prime={\rm diag}(SO_{\rm Lorentz}(D)\times SO_{\rm R}(D))~.
\eeq
The continuum twist of $\cN = 4$ that is the starting point of the twisted
lattice construction was first written down by Marcus in 1995 \cite{Marcus:1995mq}.
It now plays an important role in the Geometric Langlands program and is hence
sometimes called the GL-twist \cite{Kapustin:2006pk}.
In the case of $\cN=4$ super Yang-Mills this amounts to treating the original
four Majorana fermions as a $4\times 4$ matrix and subsequently expanding this
matrix on products of Dirac gamma matrices
\beq
\Psi=\eta I+\psi_\mu \gamma_\mu+\chi_{\mu\nu}\gamma_\mu\gamma_\nu+
\psib_\mu\gamma_5\gamma_\mu+\etab \gamma_5
\eeq
The sixteen component fields $(\eta,\psi_\mu,\chi_{\mu\nu},\psib_\mu,\etab)$
($\chi_{\mu\nu}$ is antisymmetric)
are the twisted fermions. In a similar fashion, four of the scalars which originally
transformed as a vector under the $SO(4)$ flavor subgroup
become vectors $B_\mu$ under the twisted rotational symmetry and combine
with the usual gauge fields $A_\mu$ to produce complexified gauge fields
$\cA_\mu=A_\mu+iB_\mu$ in the twisted theory. The remaining two
scalars remain as singlets under twisted rotations.

It is actually possible to pack these twisted fields into a more compact
structure by replacing the Greek index $\mu$ running from $1\ldots 4$ with
a Roman index running from $1\ldots 5$. The sixteen twisted fermions then comprise
the set $(\eta,\psi_a,\chi_{ab})$ while the bosons can be packed into five
complex gauge fields $\cA_a$. The rationale for this final change of variables
is that the twisted action can then be written in the very simple form
\beq
S = \frac{1}{g^2} \cQ \int \Tr \left(\chi_{ab}\cF_{ab} + \eta [ \cDb_a,\cD_b ] -
\frac{1}{2}\eta d\right)+S_{\rm closed}
\label{4daction}
\eeq
where $\cQ$ represents a supersymmetry transformation that transforms as a
scalar under the twisted rotation group (its appearance in the theory parallels
that of the scalar fermion $\eta$). Furthermore the original supersymmetry algebra
implies that this charge will be nilpotent with $\cQ^2=0$ so that the first term
appearing in the action Eqn.~\ref{4daction} is trivially invariant under $\cQ$
transformations. The second $\cQ$-closed term takes the form
\beq
S_{\rm closed} = -\frac{1}{8} \int \Tr \epsilon_{mnpqr} \chi_{qr} \cDb_p \chi_{mn}~.
\label{closed}
\eeq
The supersymmetric invariance of this term then relies on the Bianchi identity
\beq
\epsilon_{mnpqr}\cDb_p\cFb_{qr} = 0~.
\eeq
The nilpotent transformations associated with the scalar supersymmetry $\cQ$
are given explicitly by
\bea
\cQ\; \cA_a &=& \psi_a \nn \\
\cQ\; \psi_a &=& 0 \nn \\
\cQ\; \cAb_a &=& 0 \nn \\
\cQ\; \chi_{ab} &=& -\cFb_{ab} \nn \\
\cQ\; \eta &=& d \nn \\
\cQ\; d &=& 0, \label{BRSTsymmetry}
\eea
where the complexified field strength $F_{ab}$ is given by
\beq
\cF_{ab} = [\cD_a, \cD_b],~~~\cFb_{ab} = [\cDb_a, \cDb_b]~,
\eeq
and the complex covariant derivatives are given by
\beq
\cD_a = \partial_a + \cA_a,~~~\cDb_a = \partial_a + \cAb_a~.
\eeq

It is important to recognize that the five-dimensional look of the theory is nothing to
be afraid of; in fact, it simply reflects the fact that this
four-dimensional field theory can be viewed as the dimensional
reduction of $\cN=1$ super Yang-Mills theory in D=10 dimensions.
The five complexified gauge connections are the ten gauge fields
of that theory. In the next section we will review how easily this
picture translates into the lattice formulation.

\section{ $\cN=4$ Super Yang-Mills Theory on the Lattice\label{sec:SYMlatt}}

The prescription for discretization is actually quite natural. The
complex gauge fields are represented as Wilson gauge fields
which take their values in the {\it algebra} of a complexified
$U(N)$ gauge group \footnote{The generators are normalized as
$Tr(T^AT^B)=-\delta^{AB} $}
\beq
\cA_a(x) \rightarrow \cU_a(\vn)=\sum_{C=1}^{N^2} T^C\cU^C_a(\vn)
\eeq
Since we need five links in four dimensions we  can simply place
these Wilson link fields
on  a hypercubic lattice with an additional body diagonal
\bea
\label{eq:mu-vectores}
\hatbmu_1 &=& (1, 0, 0, 0)\nn \\
\hatbmu_2 &=& (0, 1, 0, 0)\nn \\
\hatbmu_3 &=& (0, 0, 1, 0) \\
\hatbmu_4 &=& (0, 0, 0, 1)\nn \\
\hatbmu_5 &=& (-1, -1, -1, -1)~.\nn
\eea
Thus while $\cU_a,\,a=1\ldots 4$ are associated with the usual unit vectors of
a hypercubic lattice the field $\cU_5$ is then placed on the body diagonal link.
Notice that the basis vectors sum to zero, consistent with the use of such a
linearly dependent basis.
However, it should also be clear that a more symmetrical
choice would be preferable in which the five basis vectors are treated in an entirely
equivalent manner.  A four dimensional lattice with this higher $S^5$ point
group symmetry exists and is called
the $A_4^*$ lattice. It is constructed from the set of five basis
vectors $\hatbe_a$ pointing from the center of a
four-dimensional equilateral simplex out to its vertices together with their
inverses $-\hatbe_a$. It is the four-dimensional analog of the two-dimensional
triangular lattice.
A specific basis for the $A_4^*$ lattice is given in the form of five lattice vectors
\bea
\hatbe_1 &=&  \Big(\frac{1}{\sqrt{2}}, \frac{1}{\sqrt{6}},
\frac{1}{\sqrt{12}}, \frac{1}{\sqrt{20}}\Big)\\
\hatbe_2 &=& \Big(-\frac{1}{\sqrt{2}}, \frac{1}{\sqrt{6}}, \frac{1}{\sqrt{12}},
\frac{1}{\sqrt{20}}\Big)\\
\hatbe_3 &=& \Big(0, -\frac{2}{\sqrt{6}}, \frac{1}{\sqrt{12}},
\frac{1}{\sqrt{20}}\Big)\\
\hatbe_4 &=& \Big(0, 0, -\frac{3}{\sqrt{12}}, \frac{1}{\sqrt{20}}\Big)\\
\hatbe_5 &=& \Big(0, 0, 0, -\frac{4}{\sqrt{20}}\Big)~.
\eea
The basis vectors satisfy the relations
\beq
\sum_{m=1}^{5} \hatbe_m = 0;~\hatbe_m \cdot \hatbe_n = \Big(\delta_{mn} -
\frac{1}{5}\Big);~\sum_{m=1}^{5}(\hatbe_m)_{\mu}(\hatbe_m)_{\nu} =
\delta_{\mu \nu};~\mu, \nu = 1,\cdots,4.
\eeq

It is not hard to see that the basis vectors of $A_4^*$ are a simple
deformation of the those used in the hypercubic representation and indeed
a simple Gram matrix allows one to map between the coordinates of some
field in hypercubic representation to the physical coordinates relative
to the $A_4^*$ lattice (see \cite{Catterall:2009it} for details).
Indeed for the action and other local quantities it is not necessary to
explicitly perform this mapping; the hypercubic
lattice representation furnishes a simple arena in which one can calculate
the action, check gauge invariance and carry out supersymmetry
variations without explicit reference to the $A_4^*$ lattice.
Only when we consider questions associated with rotational invariance or
space-time dependent correlation functions do we need
to map the coordinates of lattice fields into their positions relative
to the ``physical '' $A_4^*$ lattice. We stress this point because it means
that simulations can be performed in a quite standard hypercubic lattice
set-up, without concerns about details of the $A_4^*$ lattice.

The Wilson links transform
in the usual way under ordinary {\it non-complexified} $U(N)$ lattice
gauge transformations
\beq
\cU_a(\vn)\to G(\vn)\cU_a(\vn)G^\dagger(\vn+\hatbmu_a)~.
\label{gaugetransf}
\eeq
Supersymmetric invariance then precisely implies that $\psi_a(\vn)$ live
on the same links and transform identically. A local scalar fermion $\eta(\vn)$
must clearly live on a site. It transforms accordingly,
\beq
\eta(\vn)\to G(\vn)\eta(\vn)G^\dagger(\vn)~.
\eeq
In a similar fashion we place the fermionic fields $\chi_{ab}$ on new links
leading from the origin out to $\hatbmu_a+\hatbmu_b$. In the hypercubic
representation these would correspond to
links on two and three dimensional faces associated with the hypercube.
However, there is
one crucial difference from the fields $\psi_a$ - the fields $\chi_{ab}$ are
chosen with opposite orientation on these links as encoded from their
gauge transformation property:
\beq
\chi_{ab}(\vn)\to G(\vn+\hatbmu_a+\hatbmu_b)\chi_{ab}(\vn)G^\dagger(\vn)~.
\eeq
This shows how naturally the supersymmetric degrees of freedom can be
distributed on the lattice - the sixteen fermionic degrees of freedom
at a site can all be associated with the sixteen distinct links that
can be drawn in the unit four dimensional hypercube located at that site. 

To complete the discretization we need to describe how continuum derivatives are
to be replaced by difference operators. A natural technology for accomplishing
this in the case of adjoint fields was developed many years ago. It yields
expressions for the derivative operator applied to arbitrary lattice p-forms
\cite{Aratyn:1984bd}, and is thus very naturally tied to geometry. In the case
discussed here, we need just two derivatives given by the expressions
\bea
\cD^{(+)}_a f_b(\vn) &=& \cU_a(\vn)f_b(\vn + \hatbmu_a) -
f_b(\vn)\cU_a(\vn+ \hatbmu_b)~,\\
\cDb^{(-)}_a f_a(\vn) &=& f_a(\vn)\cUb_a(\vn)-
\cUb_a(\vn - \hatbmu_a)f_a(\vn - \hatbmu_a)~.
\eea
These difference operators appeared automatically as a result of orbifold projection in
the original constructions of supersymmetric lattice Yang-Mills theories from matrix models \cite{Cohen:2003xe,Cohen:2003qw,Kaplan:2005ta}.
A beautiful feature has appeared here: the construction of supersymmetric lattice
gauge theories by means of orbifolding is in one-to-one correspondence
with a simple geometrical principle \cite{Damgaard:2008pa}. Indeed,
using this geometrical prescription is by far the easiest way to see
how this lattice theory emerges.
The lattice field strength is given by the gauged forward difference
acting on the link field: $\cF_{ab}(\vn) = \cD^{(+)}_a \cU_b(\vn)$.
It is automatically antisymmetric in its indices. Furthermore,
as hoped for it transforms like a lattice 2-form and
yields a gauge invariant loop on the lattice when contracted with
$\chi_{ab}(\vn)$ (this is precisely the reason that the field $\chi$
is chosen to have opposite orientation relative to $\psi_a$).
Similarly, the covariant backward difference appearing in
$\cDb^{(-)}_a \cU_a(\vn)$ transforms as a 0-form or,
correspondingly, as a site field. It can hence can be contracted
with the site field $\eta(\vn)$ to yield a gauge invariant combination.
Thus, the twin requirements of gauge invariance and supersymmetry 
naturally places strong constraints on the whole construction.

Furthermore, this use of forward and backward difference operators guarantees that the solutions
of the lattice theory map one-to-one with the solutions of the continuum theory and
the fermion doubling problems are hence evaded \cite{Rabin:1981qj}. Another way to
understand this is to see that by
introducing a lattice with half the lattice spacing one can map this \KD fermion
action into the action for staggered fermions \cite{Banks:1982iq}. We emphasize that,
unlike the case of two-flavor or three-flavor QCD, there is no rooting problem
in this supersymmetric
construction since the additional lattice fermion degeneracy is precisely as
already required in the continuum theory.

Just like the continuum theory, the lattice action again  contains a $\cQ$-exact term:
\beq
S = \sum_{\vn} \Tr \cQ \Big(\chi_{ab}(\vn)\cD_a^{(+)}\cU_b(\vn) +
\eta(\vn) \cDb_a^{(-)}\cU_a(\vn) - \frac{1}{2}\eta(\vn) d(\vn) \Big)~.
\eeq
Acting with the $\cQ$ transformation on the lattice fields and integrating out
the auxiliary field $d$, we obtain the gauge and $\cQ$-invariant lattice action:
\beq
\label{eq:2d-latt-action}
S_0 = \sum_{\vn} \Tr \Big(\cF_{ab}^{\dagger}(\vn) \cF_{ab}(\vn) +
\frac{1}{2}\Big(\cDb_a^{(-)}\cU_a(\vn)\Big)^2 -
\chi_{ab}(\vn) \cD^{(+)}_{[a}\psi_{b]}(\vn) -
\eta(\vn) \cDb^{(-)}_a\psi_a(\vn) \Big)~.
\eeq
As in the continuum theory, the $\cQ$-exact action must be augmented by
an additional piece that is only $\cQ$-closed,
\beq
S_{closed} = \frac{1}{2} \epsilon_{abcde}
\chi_{de}(\vn + \hatbmu_a + \hatbmu_b +
\hatbmu_c) \cDb^{(-)}_{c} \chi_{ab}(\vn + \hatbmu_c),
\eeq
which is the direct analog of Eqn.~(\ref{closed}) in the continuum theory.
Remarkably, and as shown in \cite{Catterall:2007kn},
an exact lattice analog of the Bianchi identity,
\beq
\epsilon_{abcde}\cDb^{(-)}_c\cFb_{ab}(\vn + \hatbmu_c)=0,
\eeq
guarantees that the above term is invariant under $\cQ$-transformations on
the lattice. Note, incidentally, that the coefficient in front cannot be
chosen freely. Only with the specific coefficient shown will we recover
the correct naive continuum limit with full supersymmetry and Lorentz
invariance. It is intriguing to speculate what happens to this
relative coefficient 
under radiative corrections.

To show that the full action correctly reproduces the continuum theory
in the naive continuum limit, one must (in some suitable gauge) expand the gauge fields
around the unit matrix,
\beq
\cU_a(\vn)=I+a\cA_a(\vn)
\label{eq:exp}
\eeq
The following interesting phenomenon occurs:
Usually the unit matrix appearing here arises trivially once one expands
the group element $U_\mu=e^{aA_\mu}$ in powers of the
lattice spacing. However supersymmetry requires that the bosons
and the fermions be treated on an equal footing. Since the fermions are
expanded in the algebra this necessitates doing the same for the bosons.
Usually this would be a disaster since it would make it impossible
to introduce the expansion seen in Eqn.~(\ref{eq:exp}) without breaking gauge
invariance. However, in the case of a complexified $U(N)$ gauge group we
have another option: the unit matrix can arise from the
acquired vacuum expectation value of a dynamical field in the theory --
here the trace mode of the imaginary
part of the connection or, equivalently, the trace mode of the scalars in
the original (untwisted) theory\footnote{A recent construction employing only $SU(N)$ gauge
symmetry is discussed in \cite{Kanamori:2012et}.}

This expectation value can be achieved by adding to the supersymmetric action a gauge
invariant potential of the form \cite{Hanada:2010qg}
\beq
S_M = \mu_L^2 \sum_\vn \left(\frac{1}{N}{\rm Tr}
(\cU_a^\dagger(\vn) \cU_a(\vn))-1\right)^2~.
\label{eq:sm}
\eeq
Here $\mu_L$ is a tunable mass parameter, which can be used to control
the fluctuations of the lattice fields. Notice that such a potential obviously
breaks supersymmetry -- however because of the exact supersymmetry at
$\mu_L = 0$ all supersymmetry breaking counterterms induced via quantum effects
will possess couplings that vanish as $\mu_L \to 0$ and so can be removed by
sending $\mu_L\to 0$ at the end of the calculation. By adopting the polar
parametrization $\cU_a=e^{A_a+iB_a}$ it should be clear that the leading effect
of this term is to set the expectation value of the trace mode of $B_a$ to
unity as required. Furthermore, fluctuations of this trace mode are
governed by the mass $\mu_L$ while all traceless scalar modes feel only a
quartic potential. Thus the limit $\mu_L\equiv\mu a\to 0$ restores the usual
flat directions associated with the $SU(N)$ sector as the lattice spacing
$a\to 0$. A finite mass remains for the $U(1)$ mode but since this naively
decouples in the continuum limit our expectation is that
this should not lead to any observable effects in the $SU(N)$ sector.
This is one of the key issues we wish to investigate in this paper.

The above discussion illustrates the subtle way in which the
continuum limit of this theory must be reached. Without a vacuum expectation
value of the scalar trace mode, even the notion of a four dimensional continuum limit
with canonically propagating degrees of freedom cannot be introduced.

Once one has such a lattice action an obvious thing to do is to perform a
strong-coupling expansion. Normally, such an expansion around
infinitely strong bare gauge coupling reveals a phase of the theory
that is non-universal, confining, chirally broken and with a mass
gap that is given in terms of the strong coupling string tension.
Remarkably, such a standard strong-coupling expansion is not easily
implemented in this theory. It is exact supersymmetry, or rather exact
$\cQ$-symmetry that gets in the way: this theory is massless and
has only {\em one} coupling to all fields. The (inverse) bare coupling
multiplies all terms of the lattice action. This suggests that the
only consistent strong-coupling expansion will be based on
expanding the full Boltzmann factor and bringing down powers of
the action. However, there is then no damping term of the functional
integral. The bosonic degrees of freedom have non-compact support,
and the Grassmann integrals provide the heuristic 'zero' that
nevertheless could give formal meaning to such an expansion. However,
the precise way in which such an expansion scheme could be
implemented seems, at best, to be unclear. It is tempting to view
the lack of a natural strong-coupling expansion as evidence that
this theory indeed may have no strong-coupling, confining,
phase at all.

One further complication should be discussed: the potential sign problem
in the lattice theory. To simulate the theory requires carrying out
an integration over the fermions. This process generates a Pfaffian
which is generically {\it complex}. This invalidates the usual
Monte Carlo method for computing observables since the measure is no
longer real and positive definite. However, in earlier numerical work it
has been shown that the phase is actually very small for this theory,
at least after dimensional reduction to
two dimensions and contrary to the naive expectation \cite{Catterall:2011aa,Mehta:2011ud,Galvez:2012sv}.
To understand this result,
one can compute the partition function at one loop. This
was done in Ref.~\cite{Catterall:2011pd} with the result that the exact
supersymmetry leads to a perfect cancelation between bosons and fermions
and no phase appears in the final effective action. This is equivalent
to the statement that the Pfaffian is in fact real and positive definite
when evaluated on the moduli space corresponding to constant complex
commuting matrices\footnote{At least in the four supercharge case, 
this phenomenon can be related to the so-called Neuberger $0/0$ problem which presents
a hurdle to constructing a BRST transformation in lattice gauge theories where the fields are defined on a finite
group manifold. For a discussion of this connection see Ref.~\cite{Mehta:2011ud} }
Furthermore since the partition function is a topological
invariant it can be calculated {\it exactly} at one loop -- so this result
holds to all orders in perturbation theory. Since
the expectation value of the Pfaffian phase factor in the phase
quenched ensemble is proportional
to this full partition function, this argument suggests that the phase should play no role
in the lattice theory.
Of course, these arguments require exact $\cQ$ supersymmetry,
which is broken by the mass term we use to control the fluctuations of the
scalar trace mode.  Since the arguments
given above do not depend on this dimensional reduction, one could expect
it should also be true in the full four-dimensional theory.  As we will show in
the next section, our numerical results for the bosonic action (related to a derivative of the
partition function) back up this conclusion -- it approaches the exact
supersymmetric value as $\mu_L\to 0$ in the phase quenched ensemble.

One final wrinkle occurs when we contemplate doing simulations with periodic
fermion boundary conditions in all four dimensions - as is natural in
an exactly supersymmetric Euclidean theory.  The form of the fermion
action then allows for an exact zero mode of the form
$(\eta^A,\psi_a^A\chi_{ab}^A)=(\delta^{0A},0,0)$
on {\it any} background gauge field ($A$ is an adjoint index here).
This zero mode can be lifted either by use of a thermal boundary condition
or by the addition of a supersymmetric term
\beq
S_{extra}= \mu_F \cQ[\Tr(\eta)\Tr(U_aU^\dagger_a)]~.
\eeq
Performing the $\cQ$ variation leads to two new contributions to the action
\beq
\Tr[\cDb^{(-)}_a\cU_a]\Tr(\cU_a^\dagger \cU_a)-\Tr(\eta)\Tr(\psi_a U^\dagger_a)~.
\eeq
The second of these removes the fermion zero mode.
Thus  the complete action to be simulated is
\beq
S= S_0 + S_{closed} +S_M + S_{extra}.
\eeq
Although $\cQ$-exact, we should emphasize that the last piece $S_{extra}$ has
no analog in the full $\cN=4$ super Yang-Mills theory. Thus also this term must
be tuned to zero before continuum results can be extracted. In practice we have confined our study to
systems with antiperiodic boundary conditions and thsi additional term $S_{extra}$ is set to zero.

As usual in Monte Carlo simulation, the fermion variables are integrated out
and their effect in the simulation is represented by a set of  pseudofermion fields.
Notice though that the integration measure involves
only the fields $(\eta,\psi_a,\chi_{ab})$ and not their complex conjugates.
Thus it is a Pfaffian rather than a determinant that is generated. Up to a phase
this in turn
can be produced with a pseudofermion action of the form
\beq
S_{PF}=\Phi^\dagger (M^\dagger M)^{-\frac{1}{4}}\Phi\, ,
\eeq
where $M$ is the antisymmetric twisted fermion bilinear in $S$.
The fractional power of the matrix is approximated by a partial fraction
(multimass)
expansion implemented using the Rational Hybrid Monte Carlo (RHMC) algorithm.

Thus for the lattice practitioner we have a system of
\begin{itemize}
\item A set of bosonic variables appearing as noncompact complex gauge fields
\item A set of (twisted) fermions whose effect can be encoded using the
usual RHMC algorithm.
\item Both sets of fields are defined over a hypercubic lattice with additional face and body links.
\end{itemize}

This is somewhat Baroque, but it is simple and it is completely manageable.
It might be useful to list what can be computed at this stage:

\begin{itemize}
\item
We can simulate with
periodic or
antiperiodic fermionic boundary conditions so that we can do either
zero or finite temperature (supersymmetry-breaking) simulations
\item
We can dial in various masses ($\mu_L$, $\mu_F$) to explicitly break
various symmetries. This will ultimately
be useful for computing critical exponents.
\item
We can compute eigenvalues of the fermion (Dirac) operator.
\item We can measure Wilson and Polyakov lines to extract, for example,
the static quark-antiquark potential and look for confinement/deconfinement.
\item We can monitor the distribution of gauge invariant scalar eigenvalues
extracted from the observable $\cU_a^\dagger \cU_a$ which gives us a handle on
possible problems associated with integration over the flat directions.
\end{itemize}

Finally, we should stress the following. To obtain physical correlation
functions from this twisted theory, one must perform the appropriate
un-twisting on observables. In terms of our twisted variables, physical
quantities will generically appear in rather complicated combinations
of the variables described here. However, the map is straightforward
and can easily be implemented in measurements.
And, in the cases of spectral observables, we do not need to perform the un-twisting:
operators with the same sets of space-time symmetries couple to the same
set of physical states; only the
relative coupling coefficients will be different.

\section{Simulation Results}
\label{sec:sim-results}

\subsection{Introduction to the simulations\label{sec:sim-intro}}

We begin with a few words about lattice observables. As usual,
gauge invariance implies
quite strict limitations on the observables we can construct out of
our lattice variables. What is new in this theory as compared to
ordinary lattice gauge theory is the natural appearance of link
variables that live in the algebra of the gauge group rather than
in the group itself. 
This also
implies that the integration measure naturally is over anti-Hermitian
gauge variables rather than being the invariant gauge group (Haar) measure.
The reason is that the measure must remain invariant under an arbitrary
shift symmetry, as is clear from Eqn.~\ref{BRSTsymmetry}. This brings
to the open an important point regarding the ordinary Yang-Mills
gauge symmetry of this theory: The gauge transformations of the gauge
links (which live in the algebra of the gauge group) are defined
by the multiplication rule (\ref{gaugetransf}). At first glance it is not obvious
that the flat integration measure associated with the gauge links is invariant
under these gauge transformations: the non-linear transformation
begs for the left and right invariant Haar measure instead. However, since the
links are complexified one must integrate over both the field and its complex conjugate and
this saves the day; the Jacobians arising after a gauge transformation
cancelling against each other leaving the final measure invariant
as required\footnote{We thank Issaku Kanamori for pointing this out}.
However this argument fails for the fermion link fields since they do not appear with
their complex conjugates in the measure. Remarkably, however
one does find that the ordinary flat
measure {\em is} invariant after taking the product over all lattice
points. This is not totally surprising from the point of view of
the orbifolding construction, and it is instructive to see how it
arises in detail.

The gauge transformation for a typical fermion
link variable such as $\psi_a(\vn)$ is written in eq. (\ref{gaugetransf}),
\beq
\psi_a(\vn)\to G(\vn)\psi_a(\vn)G^\dagger(\vn+\hatbmu_a)\, ,
\eeq
where $\psi_a(\vn) = T^A\psi^A_a(\vn)$ and we integrate over
the flat measure in the variables $\psi^A_a(\vn)$.

On a finite lattice, $G^\dagger(\vn+\hatbmu_a)$ is a function
different from $G(\vn)$. It is still sufficient to check
gauge invariance for infinitesimal (but different)
transformations. Let us choose
\begin{eqnarray}
           G(\vn) &=& 1 + \alpha^A T^A \cr
           G(\vn+\hatbmu_a) &=& 1 + \beta^A T^A ~.
\end{eqnarray}
Expanding and collecting terms we get the same cancellations as
in the continuum plus two new terms in the transformation law
for $\psi_a(\vn)$:
\beq T^A \psi^A_a(\vn)\to \alpha^A T^A T^B \psi^A_a(\vn)
-  \beta^A T^A T^B \psi^A_a(\vn)
\eeq
In the naive continuum limit,
where $\alpha^A - \beta^A \sim a$, these terms can be ignored
and the usual gauge invariance of the continuum is recovered.
But for finite lattice spacing $a$ the new terms remain. However,
on the group $U(N)$ we can always expand a product $T^A T^B$ in
the generators of the group:
\beq
          T^A T^B  =  \frac{i}{2}f^{ABC} T^C  + d^{ABC} T^C\, ,
\eeq
where $d^{ABC}$ are the symmetric structure constants. We can now read off
the additional terms in the transformation of the components $\psi^A_a$.
The first new piece vanishes because
of $f^{AAC} = 0$, and only the second piece remains. For the
link fermion $\psi_a(\vn)$ it is of the form $(\alpha - \beta)^C d^{AAC}$,
which does not vanish. However, the measure is the product
$\prod d\psi_a^A$ over all links on the lattice. Link for link
the leftover pieces cancel among each other because of the conjugation
involved in the gauge transformation (\ref{gaugetransf}). It is
interesting to see how gauge invariance is not insured for a single
link, but recovered once the transformations of the neighboring
links are included. From the orbifolding construction one could
perhaps have guessed that such a mechanism would need to be invoked.

It should be noted here that even for conventional U(1) lattice gauge theories decompactified gauge-fields 
obtained via stereographic projection of the group manifold can be used in order to construct a lattice BRST symmetry,
see \cite{Mehta:2009,vonSmekal:2008es,vonSmekal:2007ns}.

The choice of gauge group is clearly not very essential for a
first set of simulations. For simplicity,
we have here simulated the (phase quenched) $U(2)$ theory on lattices of size
$L=4^4,6^4, 8^4$  for a wide range of bare `t Hooft
couplings $\lambda=0.2-2.6$ and values of the regulator mass in the range
$\mu_L=0.1 - 1.0$. The simulations have mostly been performed using anti-periodic
(thermal) boundary conditions for the fermions. This evidently breaks
supersymmetry, but it also removes an exact fermionic zero momentum mode
associated the trace mode of fermions that is otherwise present. The
breaking due to anti-periodic boundary conditions turns out to be
tiny, and will of course disappear as larger volumes are being considered.

An RHMC algorithm has been used for the simulations. It has been described
in detail in ref. \cite{Catterall:2011cea}. The use of a GPU accelerated
solver \cite{Galvez:2011cd} has allowed us to reach larger lattices than
have thus far been studied. It is important to recognize that the supersymmetric fermion operator
defined on a lattice of size $L$ is equivalent, in terms of counting degrees of freedom, to a staggered operator on
a lattice of size $2L$.

\subsection{Lattice moduli stabilization\label{sec:sim-eigenval}}

As in continuum $\cN = 4$ super Yang-Mills theory, the lattice theory possesses flat
directions corresponding to the continuum of
classical vacuum states in which the bosonic fields take values in the space of
constant, mutually commuting, complex matrices. This continuum of
vacuum states is called the moduli space of the theory and is determined by the expectation values of
the scalar fields appearing as imaginary parts of gauge fields in the
twisted formulation.
Potential divergences appear in the partition function of the theory when integrating over
these flat directions.

In the lattice theory we stabilize these moduli
by the addition of the term Eq.~\ref{eq:sm} to the action. This term certainly lifts the moduli space of
the theory but in the lattice theory with $U(N)$ gauge symmetry it plays an even more important role
by generating a gauge invariant vacuum expectation value for the complexified Wilson link field
${\Tr}\cU_a^{\dagger} U_a=1$. This allows one to argue that in the limit $a\to 0$ and
in a fixed gauge $\cU_a\sim I+\cA_a+\ldots$. This latter expansion is {\it required} if the
naive continuum limit of the lattice theory is to target a four dimensional field theory. Furthermore, the unit
matrix that appears in this expression can then be
interpreted as corresponding to giving a fixed vev to the trace mode of the
scalar field. However, it is not clear that this vev survives quantum corrections and in principle
one needs to check this in the simulations.

\begin{figure}
\begin{center}\includegraphics[width=0.75\textwidth]{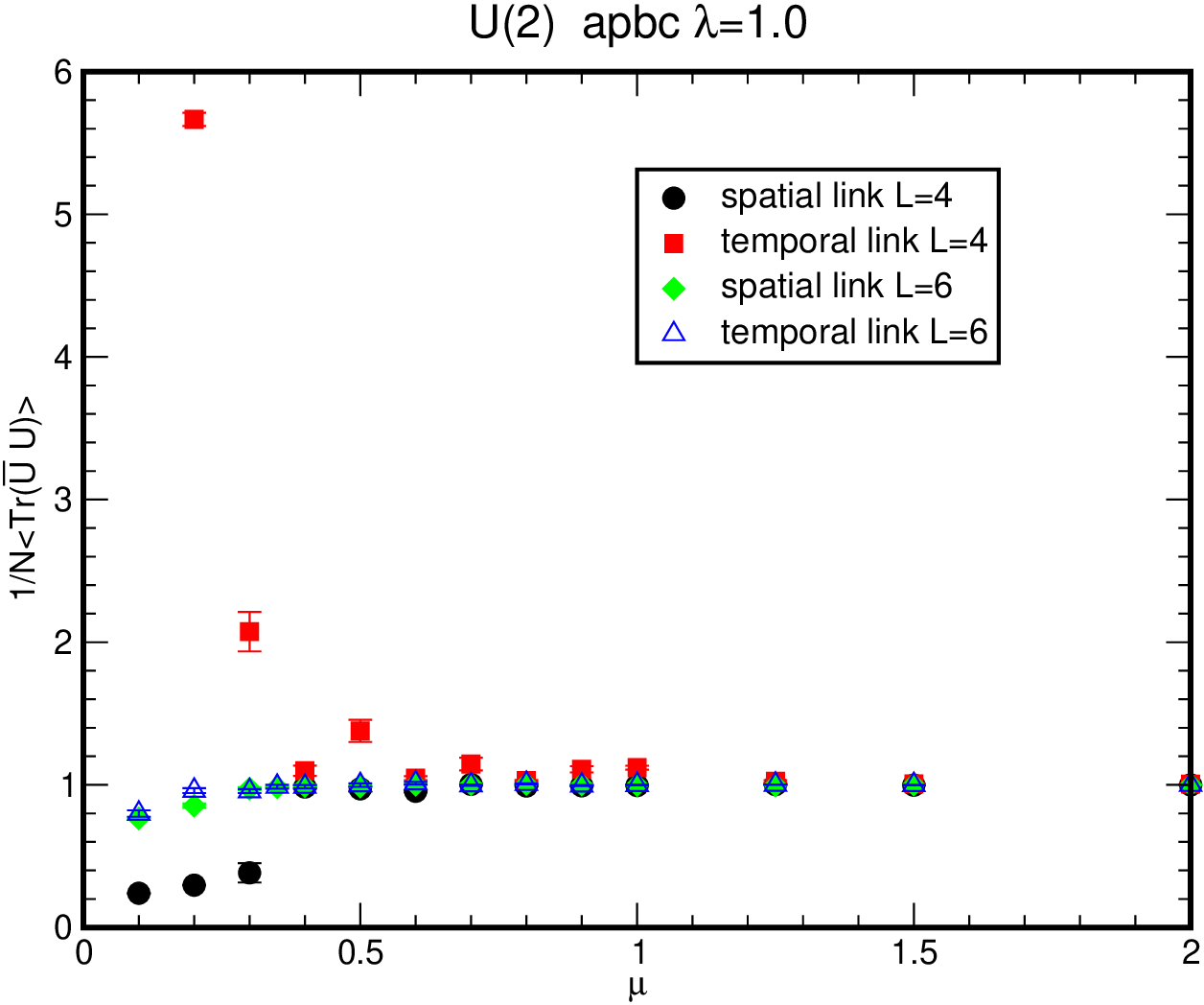}\end{center}
\caption{\label{fig:scalars}$\frac{1}{N}{\Tr}(U^\dagger_a U_a)$ vs
 mass parameter $\mu_L$ at 't Hooft coupling $\lambda=1.0$.}

\end{figure}
Clearly the correct vacuum state is picked out uniquely as $\mu_L\to\infty$. But the supersymmetric limit lies in the
opposite direction where $\mu_L\to 0$. It is important to be able to locate which regions in the bare parameter
space are consistent with such a link expectation value and simultaneously possess small
supersymmetry breaking. Figure.~\ref{fig:scalars} shows a plot of the spatial link and temporal
link expectation values versus $\mu_L$ at 't Hooft coupling
$\lambda=1.0$ on lattices of size $L=4$ and $L=6$. For small enough
$\mu_L$ the link vev can become destabilized either running to zero or large values. In such regions of the
bare parameter space we claim that there is no possible four dimensional continuum limit.
Our data indicates that this region of instability is
pushed to smaller values of $\mu_L$ for larger lattices so it is likely a finite size artifact.
All of the data we show in the following sections corresponds to regions of the phase diagram
where the link vacuum expectation value is close to unity.

\begin{figure}
\begin{center}\includegraphics[width=0.75\textwidth]{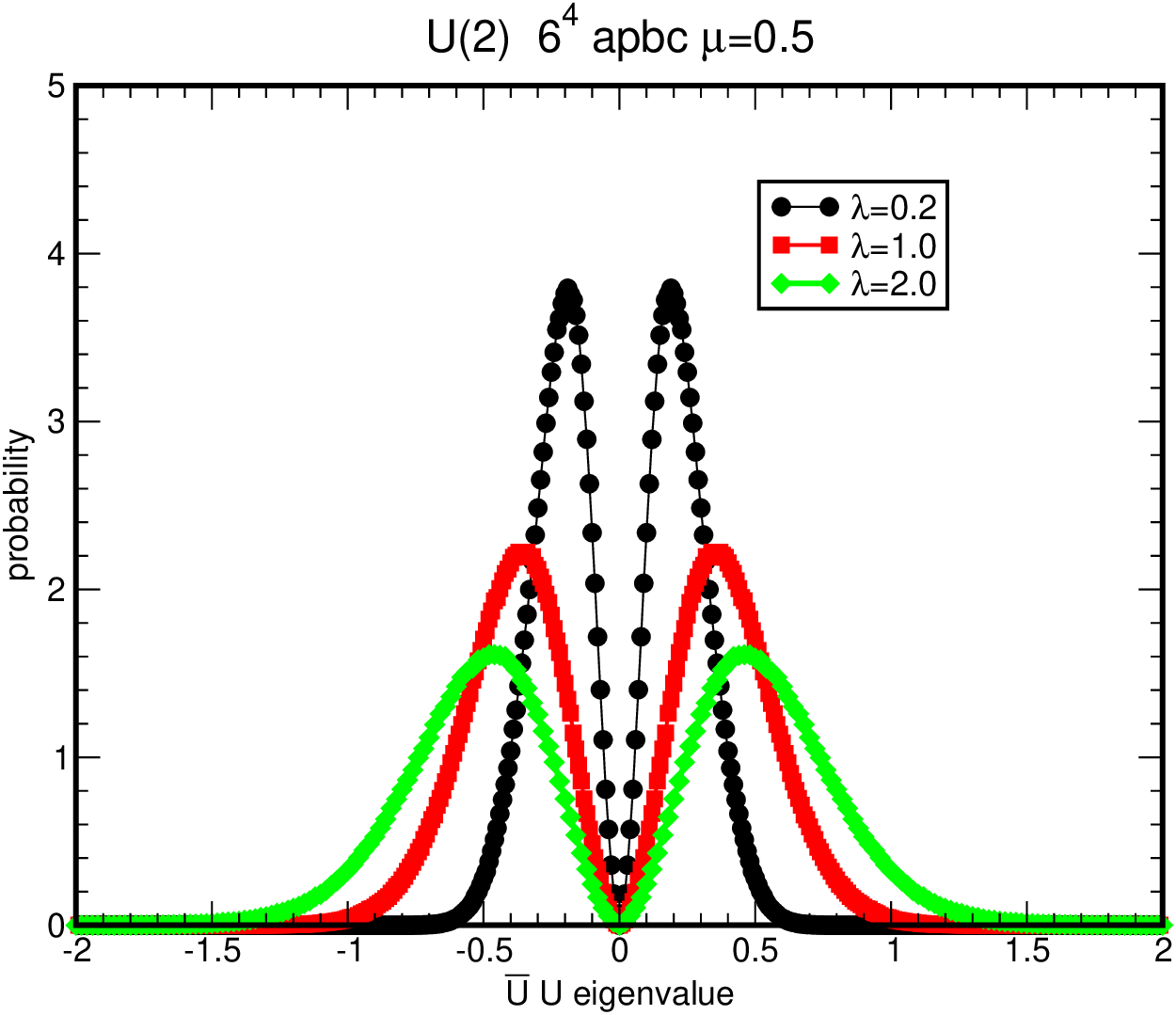}\end{center}
\caption{\label{fig:eigmu05}Eigenvalues of the traceless part of $\cU_a^\dagger \cU_a$ averaged over the Monte Carlo ensemble for $\mu_L=0.5$ and $\lambda=0.2,1.0,2.0$.}
\end{figure}

\begin{figure}
\begin{center}\includegraphics[width=0.75\textwidth]{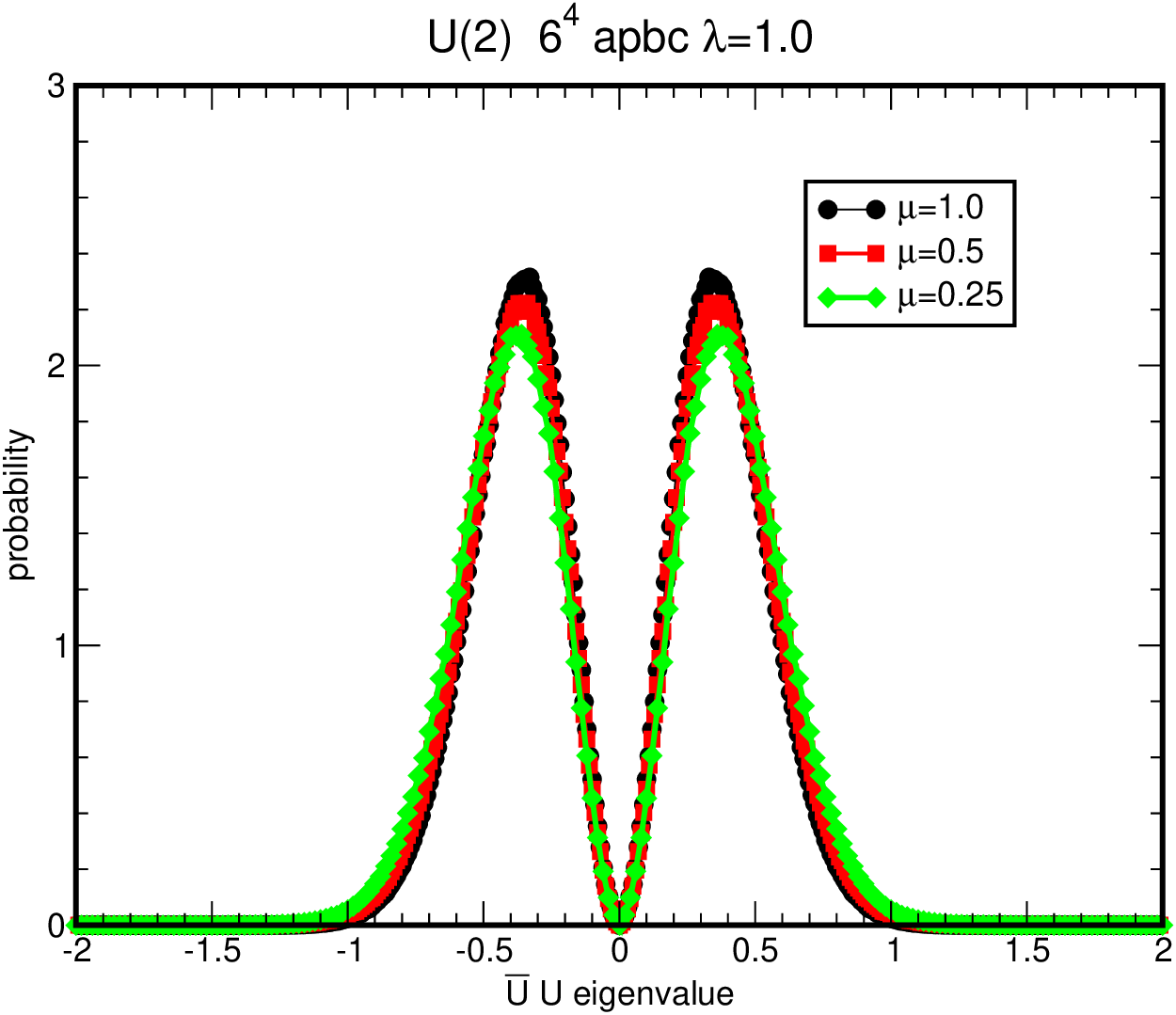}\end{center}
\caption{\label{fig:eiglam1}Eigenvalues of the traceless part of $\cU_a^\dagger \cU_a$ averaged over the Monte Carlo ensemble for $\lambda=1.0$ and $\mu_L=0.25,0.5,1.0$.}
\end{figure}

Beyond leading order, the added potential term also lifts and stabilizes
the regular $SU(N)$ flat directions and
one might also worry that as $\mu_L\to 0$ this stabilization mechanism would also prove ineffective.
To see that it does not consider the distribution of eigenvalues of $\cU_a^\dagger \cU_a$  for several
values of $\mu_L$ on a $L=6$ lattice.  Fig.~\ref{fig:eigmu05} shows the distribution of
the eigenvalues of the traceless part of this quantity for several 't Hooft couplings $\lambda$ and for
$\mu_L=0.5$. For this value of $\mu_L$ the scalar eigenvalues do not wander down the flat directions but remain
localized close to the origin in field space. The width of the resulting distributions does however
increase as the gauge coupling is increased. Of course the most interesting issue is whether
the scalar fields remain bounded as we send the supersymmetry breaking mass term to zero.
The answer seems to be in the affirmative; Fig.~\ref{fig:eiglam1} shows the distributions for
fixed 't Hooft coupling $\lambda=1.0$ as the mass parameter $\mu_L$ is decreased. The plots show a very weak
dependence on $\mu_L$ consistent with the distributions approaching a well defined limit as $\mu_L\to 0$.
However, it is important to note that this limit must be performed carefully; as we
have seen we should send $L\to\infty$ {\it before} we
can truly set $\mu_L$ to zero. If we don't do this we will encounter instabilities associated with
the flat directions. 

At first sight the apparent localization of the scalar eigenvalues close to the origin seems to
indicate that the classical moduli space is in fact lifted by quantum corrections.  Such a conclusion
would disagree with the perturbative calculation carried out in \cite{Catterall:2011pd} which shows that the single
exact supersymmetry is sufficient to ensure that the
effective potential {\it in the lattice theory} vanishes to all orders in perturbation theory in a fashion
analogous to the continuum. 

We thus do not believe that
this is the correct interpretation of the results but instead that the observed localization is 
connected to the treatment of the zero modes in the theory. First notice that the Pfaffian vanishes on the
flat directions since in the presence of a constant commuting bosonic background there appear
exact fermion zero modes. In the full path integral these would formally cancel against the corresponding bosonic
zero modes corresponding to fluctuations in the flat directions. However the supersymmetry breaking
breaking potential we have added lifts these bosonic zero modes. The net effect is that the configurations corresponding to the exact flat directions {\it do
not contribute} to the lattice path integral. Furthermore, since the valleys corresponding to the flat directions
possess increasingly steep sides as we move away from the origin in field space we expect
that the contribution of field configurations corresponding to fluctuations
away from the flat directions will yield a distribution 
in the scalar eigenvalues that has a peak close to the origin - as we observe. These effects have been
observed before by Staudacher at al \cite{Krauth:1999qw} in the context of supersymmetric matrix models.
We think that this is the correct interpretation of our eigenvalue distributions too - the zero mode
sector of $\cN=4$ on a finite lattice corresponding to the corresponding supersymmetric matrix model.

\subsection{Bosonic Action and Polyakov lines\label{sec:sim-global}}

\begin{figure}
\begin{center}\includegraphics[width=0.75\textwidth]{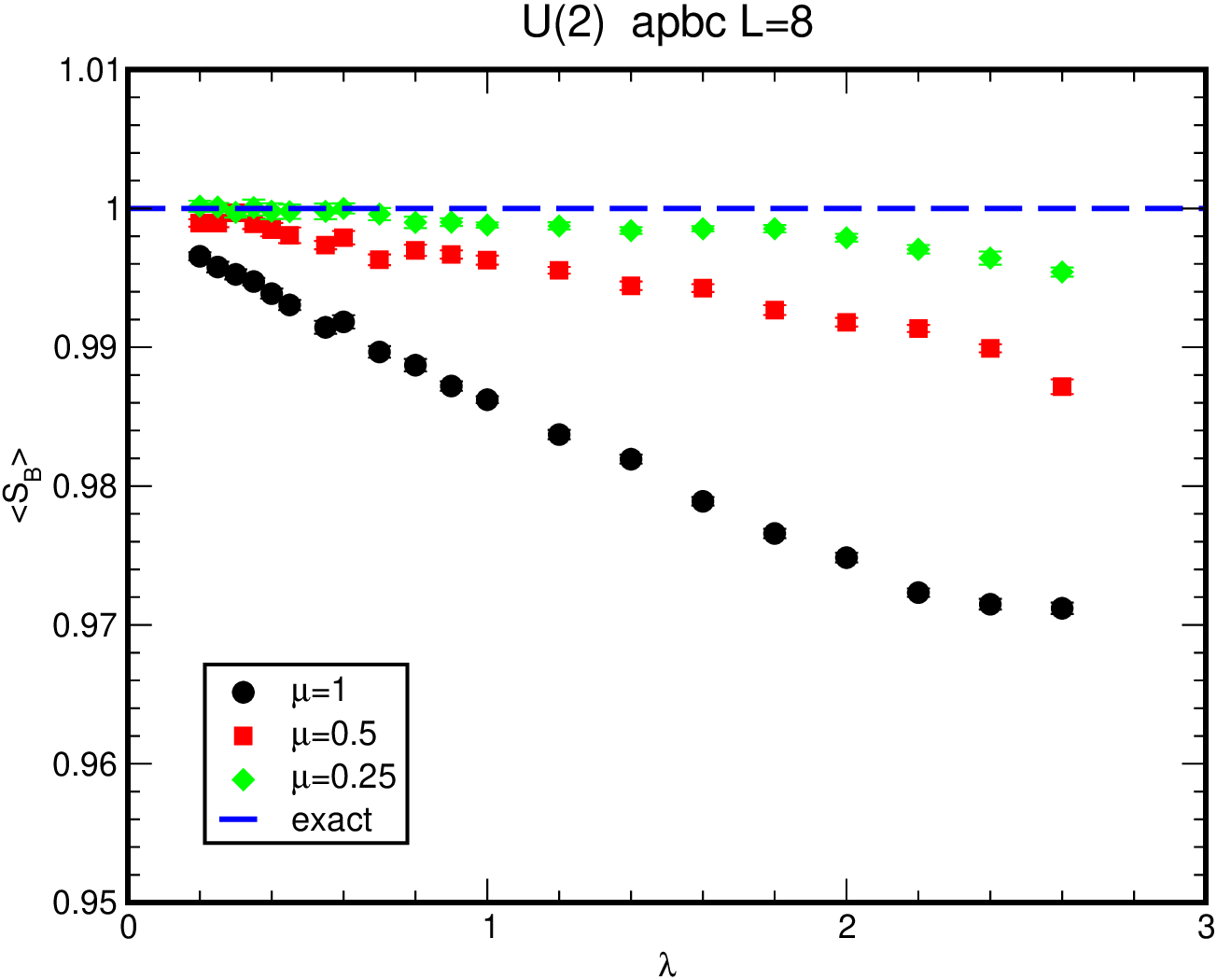}\end{center}
\caption{\label{fig:bact2}Expectation value of the bosonic action vs 't Hooft coupling $\lambda$ for
$\mu_L=0.25,0.5,1.0$. The data is normalized so that the supersymmetric result is unity.}
\end{figure}

In this initial study we have focused on understanding of the
phase diagram of the lattice theory. First let us examine the bosonic action.
This quantity is related to
$\frac{\partial\ln{Z}}{\partial\lambda}$, which vanishes on account of the
topological character of the partition function in the supersymmetric
limit\footnote{The fermions appear quadratically in the action and hence their
expectation value can be computed via a simple scaling argument}. We see in
fig.~\ref{fig:bact2}  that the
measured value for $<S_B>$ is indeed approximately $\lambda$-independent for small $\mu_L$ and agrees
very well with the exact value $S_B/(9L^4N^2/2)=1$.
Notice that this result is both consistent with exact supersymmetry in the
lattice theory and additionally lends strength to the claim that
a genuine sign problem is absent in this theory, and that the phase quenched
ensemble hence is adequate for studying the theory.

\begin{figure}
\begin{center}\includegraphics[width=0.75\textwidth]{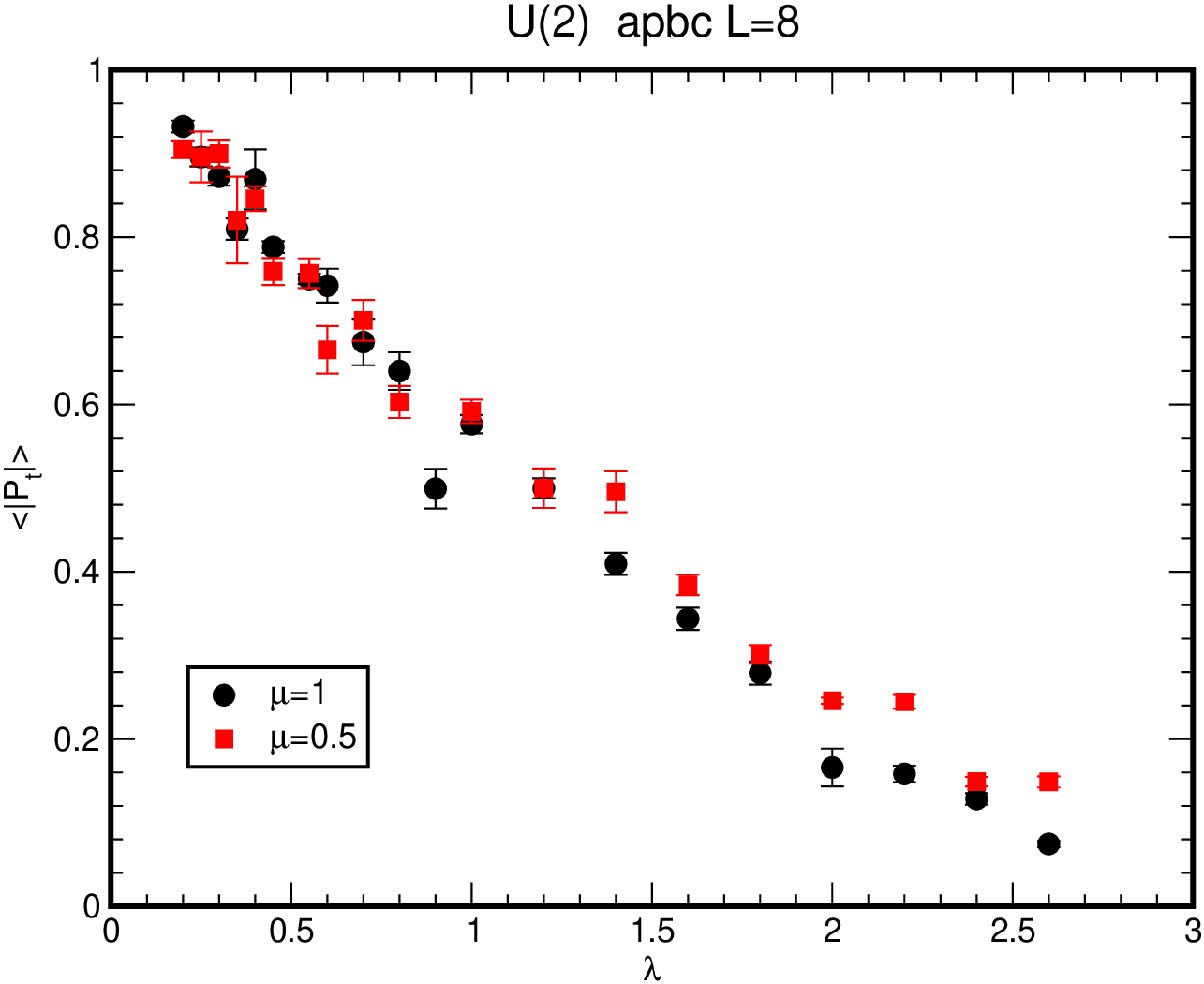}\end{center}
\caption{\label{fig:lines_t}Absolute value of the temporal Polyakov line vs $\lambda$ for
$\mu_L=1.0,0.5$ on a lattice of size $L=8$.}
\end{figure}

We now turn to the Polyakov lines. Since $\cQ \cU_a=0$ we expect that
the Polyakov line is both gauge invariant and supersymmetric. Indeed as for
the bosonic action the latter would guarantee that the Polyakov line would
take a value which was independent of $\lambda$ in the limit $\mu_L\to 0$.
Figure~\ref{fig:lines_t}
shows the (absolute value of the) temporal Polyakov line versus
bare coupling for $L=8$ and $\mu_L=0.5,1.0$. The spatial line agrees with the temporal line
within statistical errors. Unlike the bosonic action we see a dependence on
the coupling $\lambda$ and little indication that taking $\mu_L$ to zero will regain the
supersymmetric result.

\begin{figure}
\begin{center}\includegraphics[width=0.75\textwidth]{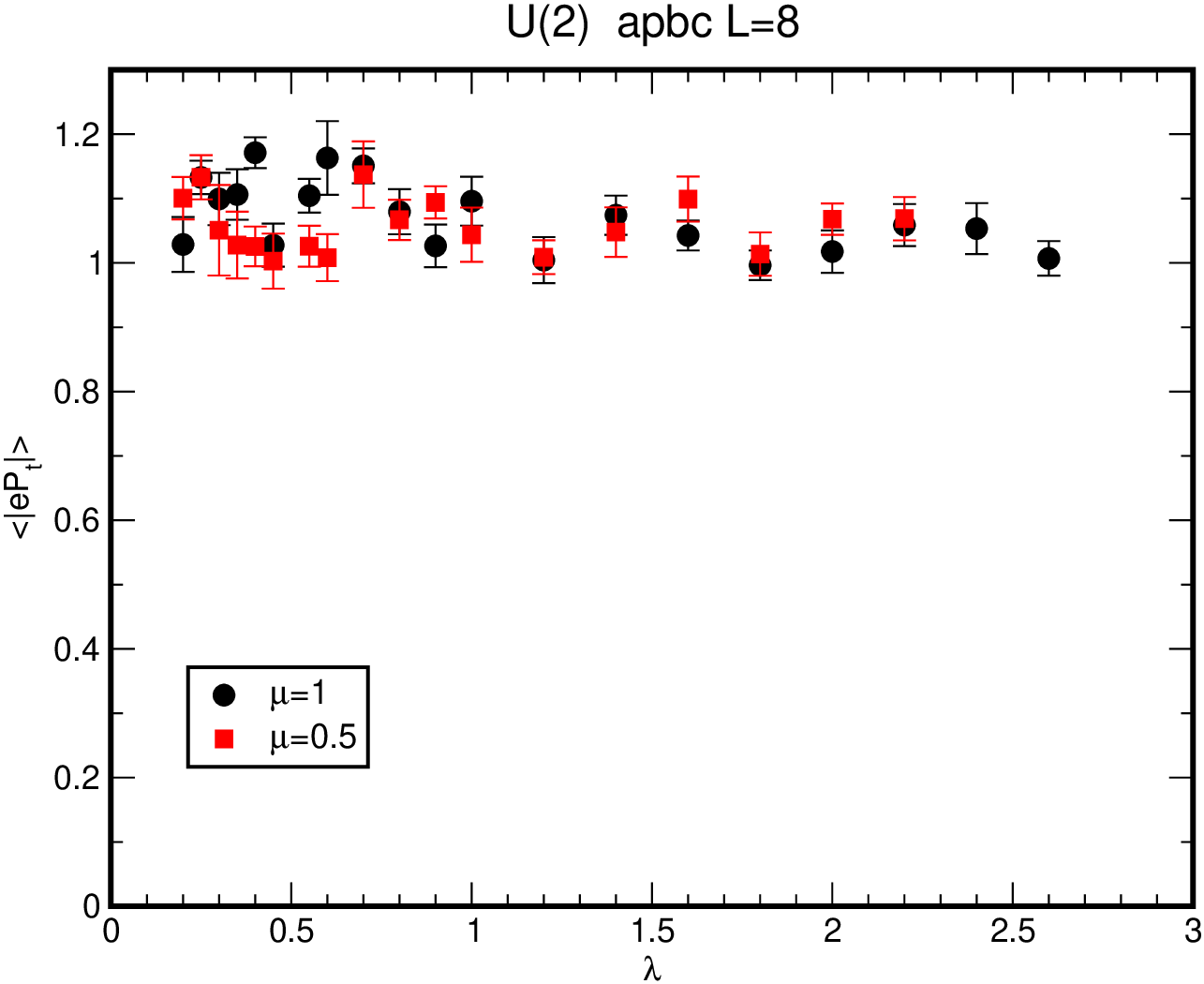}\end{center}
\caption{\label{fig:elines_t}Absolute value of the traceless part of the temporal Polyakov line vs $\lambda$ for
$\mu_L=1.0,0.5$ on a lattice of size $L=8$.}
\end{figure}

Insight into this problem can be gained by plotting a related quantity; the Polyakov line projected
to the traceless $SU(2)$ sector. This is easily accomplished by taking the traceless part of
$\cU_a(x)$ and exponentiating the result to achieve a matrix in $SL(2,C)$. The corresponding temporal
Polyakov line computed from this link is shown in fig.~\ref{fig:elines_t}.

In this case the value of the line is approximately independent of coupling $\lambda$ as one
would expect for an observable invariant under the exact supersymmetry. We deduce that
the supersymmetry breaking we are seeing is associated with the U(1) sector.
Perhaps one should not be too surprised by this; after all the potential term  we add to stabilize the
moduli space gives an explicit mass to the U(1) scalars and hence supplies a strong source of
supersymmetry breaking in this sector. Intriguingly we have also computed the Polyakov line from
the unitary projection of $\cU_a$ and find a behavior similar to that in fig.~\ref{fig:lines_t}. This is
evidence that the breaking is actually associated not with the trace mode of the scalars but
the additional massless U(1) gauge field that appears in the theory.

Let us make a final comment. Both the bosonic action and Polyakov lines show only smooth
behavior as we scan in the 't Hooft coupling even as $\mu_L\to 0$. Over the entire range we have explored,
$\lambda\le 2.6$, there are no hints of
phase transitions in the system associated with a two phase structure as one might have
naively expected. We will present  additional evidence in favor of a single lattice phase in
the next section.

\subsection{Wilson loops and the static potential\label{sec:sim-potential}}

\begin{figure}
\begin{center}\includegraphics[width=0.75\textwidth]{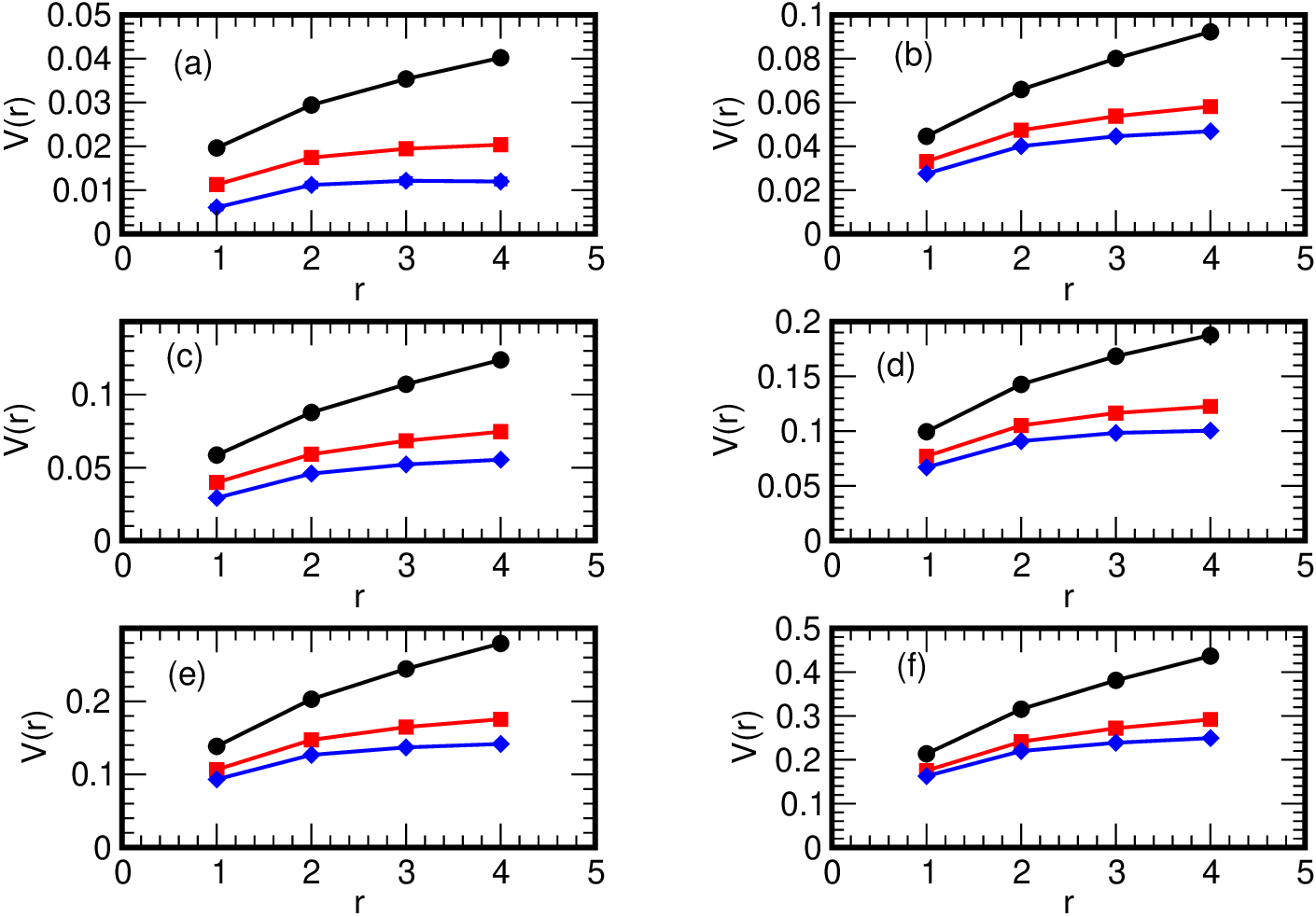}\end{center}
\caption{\label{fig:vr} Potentials from Wilson loops, from $8^4$ $\mu_L=1$ simulations.
Octagons label potentials from $t=1-2$, squares from $t=2-3$ and diamonds from $t=3-4$.
(a) $\lambda=0.25$;
(b) $\lambda=0.45$;
(c) $\lambda=0.6$;
(d) $\lambda=0.9$;
(e) $\lambda=1.2$;
(f) $\lambda=1.6$.
}
\end{figure}

Finally we turn to the static potential which we compute using the ``supersymmetric''
Wilson loops $W(L,M)$ which include the six scalars.
Denoting such Wilson loops by
$W(r,t)$ where the second index indicates that we align the loop along the
temporal direction, we can define the potential $V(r)$ to be
\beq
W(r,t)=\exp(-V(r)t)
\label{eq:wr}
\eeq
or, equivalently, we make an ``effective mass'' determination of the potential from
\beq
V(r) = -\log \frac{W(r,t+1)}{W(r,t)}.
\label{eq:vreff}
\eeq
This is a standard technique from the point of lattice QCD simulations.
Examples of this analysis from our $8^4$ data sets are shown in Fig.~\ref{fig:vr}.
The fact that the data from different $t$ values are not coincident is a sign that
$t$ is not large enough that Eq.~\ref{eq:wr} is obtained; higher-energy excitations
of the Wilson loop still contribute to $W(r,t)$. Nevertheless, the figures already
indicate that  the potential flattens to a constant at large $r$.

\begin{figure}
\begin{center}\includegraphics[width=0.75\textwidth]{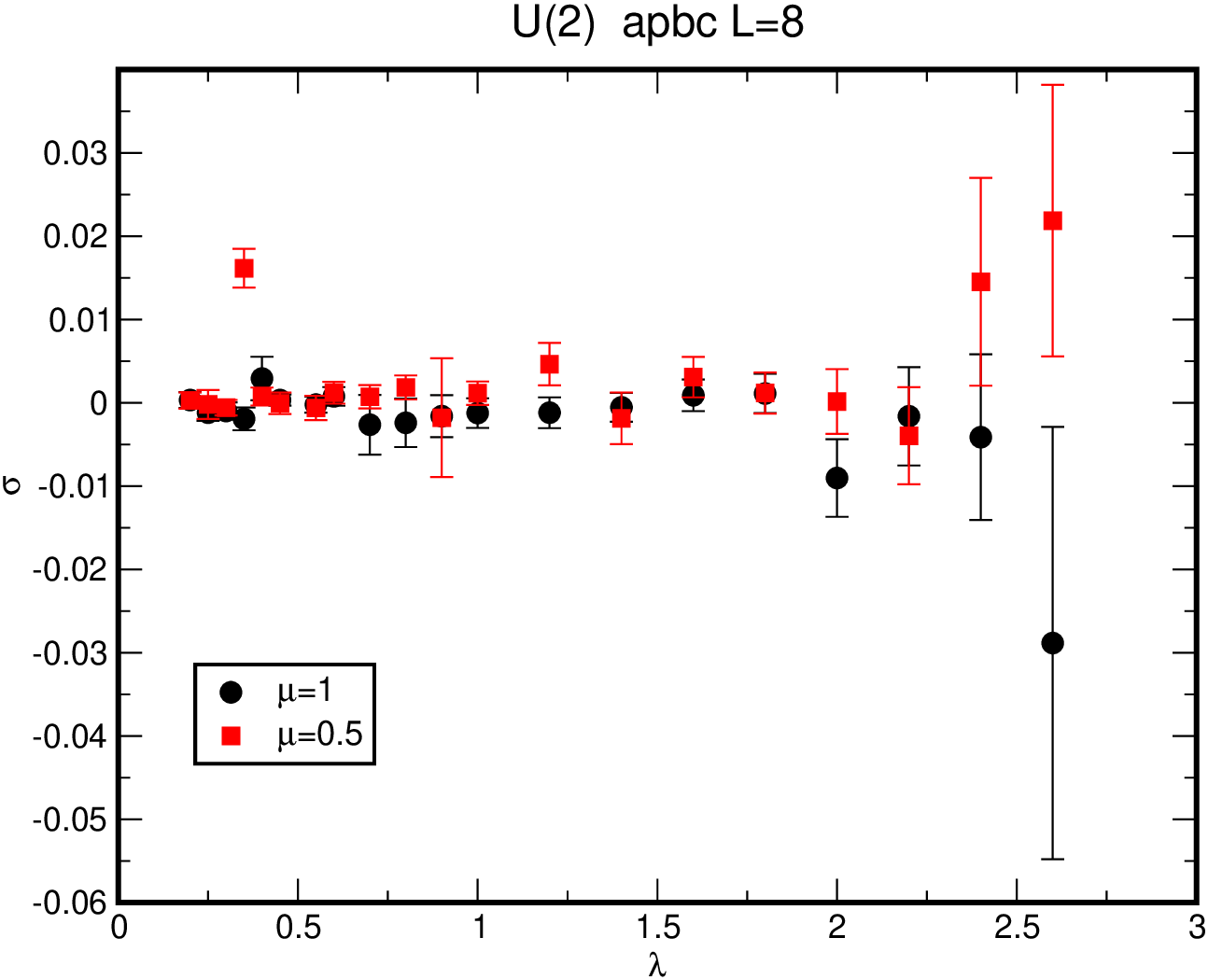}\end{center}
\caption{\label{fig:sigma}String tension from fits to
from Wilson loops with $t=3$ and 4 from $8^4$ data sets. Circles are $\mu_L=1.0$; diamonds, $\mu_L=0.5$.
}
\end{figure}

\begin{figure}
\begin{center}\includegraphics[width=0.75\textwidth]{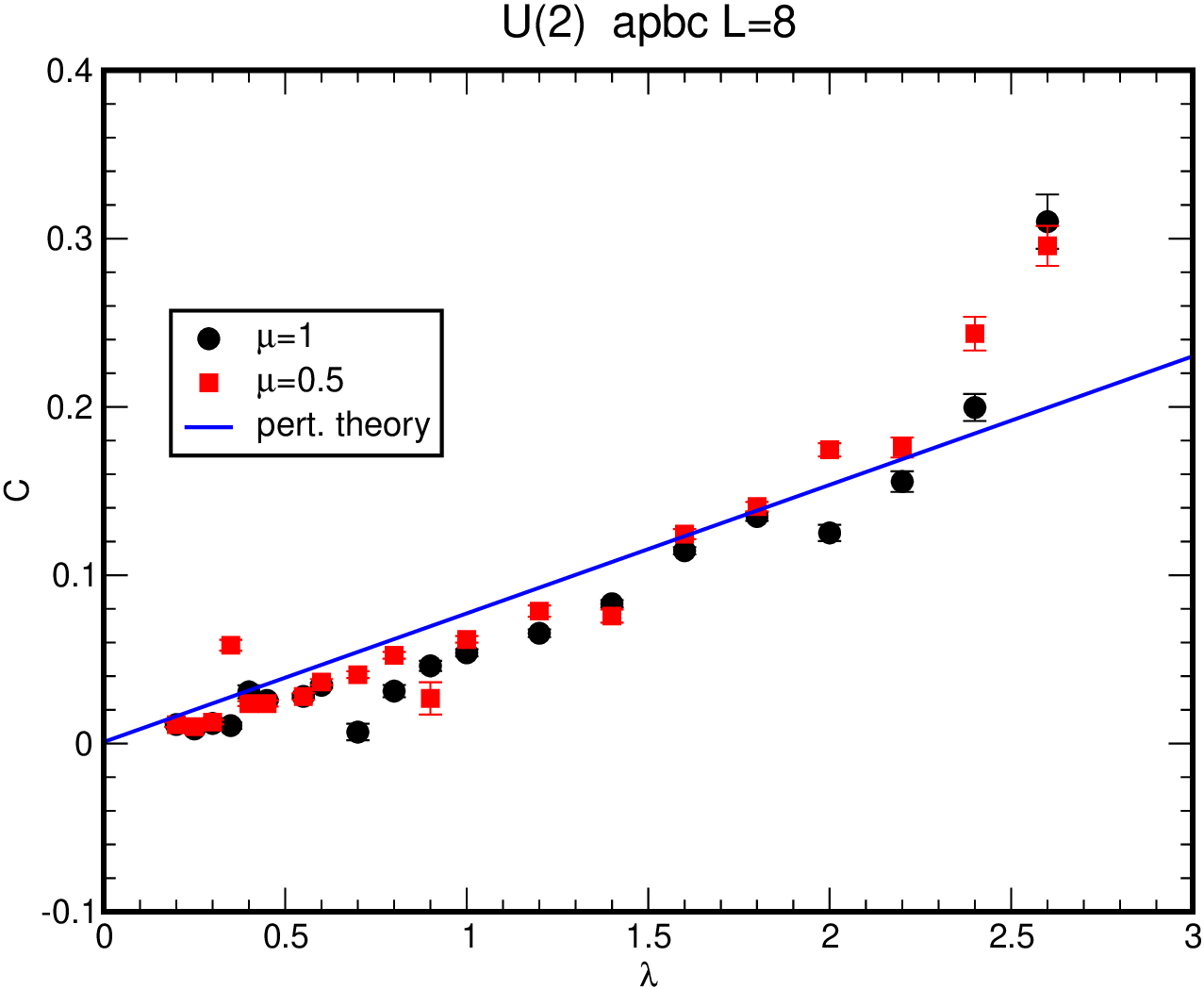}\end{center}
\caption{\label{fig:coul}Coulomb coefficient
from Wilson loops with $t=3$ and 4 from $8^4$ data sets. Circles are $\mu_L=1.0$; diamonds, $\mu_L=0.5$.
}
\end{figure}

We can make this statement a bit more quantitative by taking the largest-$t$ data
(Wilson loops at $t=3$ and 4), extracting the potential by fitting Eq.~\ref{eq:vreff},
and performing a fit to
\beq
V(r)= -\frac{C}{r} + A + \sigma r.
\label{eq:vr}
\eeq
For our data sets, with four values of $r$, we have one degree of freedom.
We observe that $\sigma\simeq 0$ and that the fits uniformly have a
$\chi^2/DoF$ smaller than unity. Of course, the quantities in the fit are highly
correlated since they come from the same underlying configurations.
Therefore, we fold the whole fit into a jackknife.
The extracted string tension $\sigma$ is shown in Fig.~\ref{fig:sigma}. Again,
it is clearly consistent with zero.
Observing zero string tension raises the possibility that $V(r)$ is, in fact,
Coulombic. We thus repeat the fit, but this time with $V(r)=A+ C/r$. Again over
the observed range of couplings we have good fits with
$\chi^2/DoF$ again less than unity, now for
two degrees of freedom. Fig.~\ref{fig:coul} shows the coefficient of the
Coulomb term as a function of the `t Hooft coupling. It is remarkably linear.
The naive expectation of perturbation theory (one gauge boson exchange) is
\beq
C=\frac{g^2N}{4\pi}~.
\eeq
This seems to describe the data well, and suggests that the strong coupling regime is
above $\lambda \ge 2.5$.

Thus the Wilson loop analysis lends support to the hypothesis of a single phase structure with
vanishing string tension for all bare couplings $\lambda$.

\subsection{Fermion eigenvalues and chiral symmetry breaking\label{sec:sim-chsb}}
\begin{figure}
\begin{center}\includegraphics[width=0.75\textwidth]{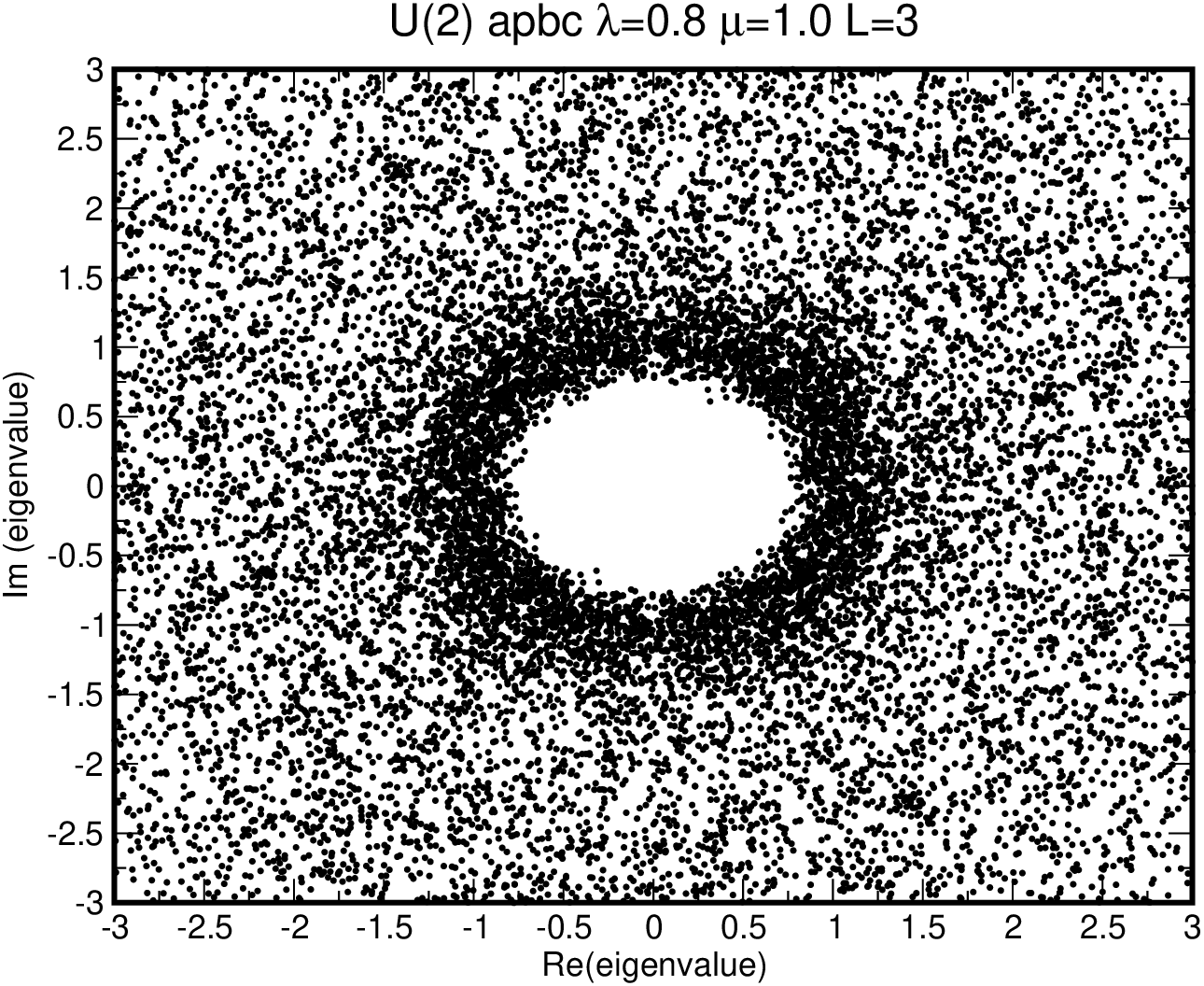}\end{center}
\caption{\label{fig:feigen}Fermion eigenvalues obtained from a Monte Carlo ensemble at $\lambda=0.8$,
$\mu_L=1$ and $L=3$.}
\end{figure}

If the system is really conformal for all gauge couplings then it should not support a chiral
condensate. To understand this better we have studied the spectrum of the twisted
fermion operator on a small lattice.
Fig.~\ref{fig:feigen} shows a scatter plot of the fermion eigenvalues coming from a run
with $\lambda=0.8$ and $\mu_L=1$ on a small $L=3$ lattice. The most obvious feature
is that {\it no} eigenvalues are found close to the origin. This is a robust statement; at all
couplings$\lambda$ a gap appears in the spectrum independent of $\mu_L$.
This, by virtue of the Banks-Casher theorem, means that chiral symmetry
is not spontaneously broken in this lattice theory.

\begin{figure}
\begin{center}\includegraphics[width=0.75\textwidth]{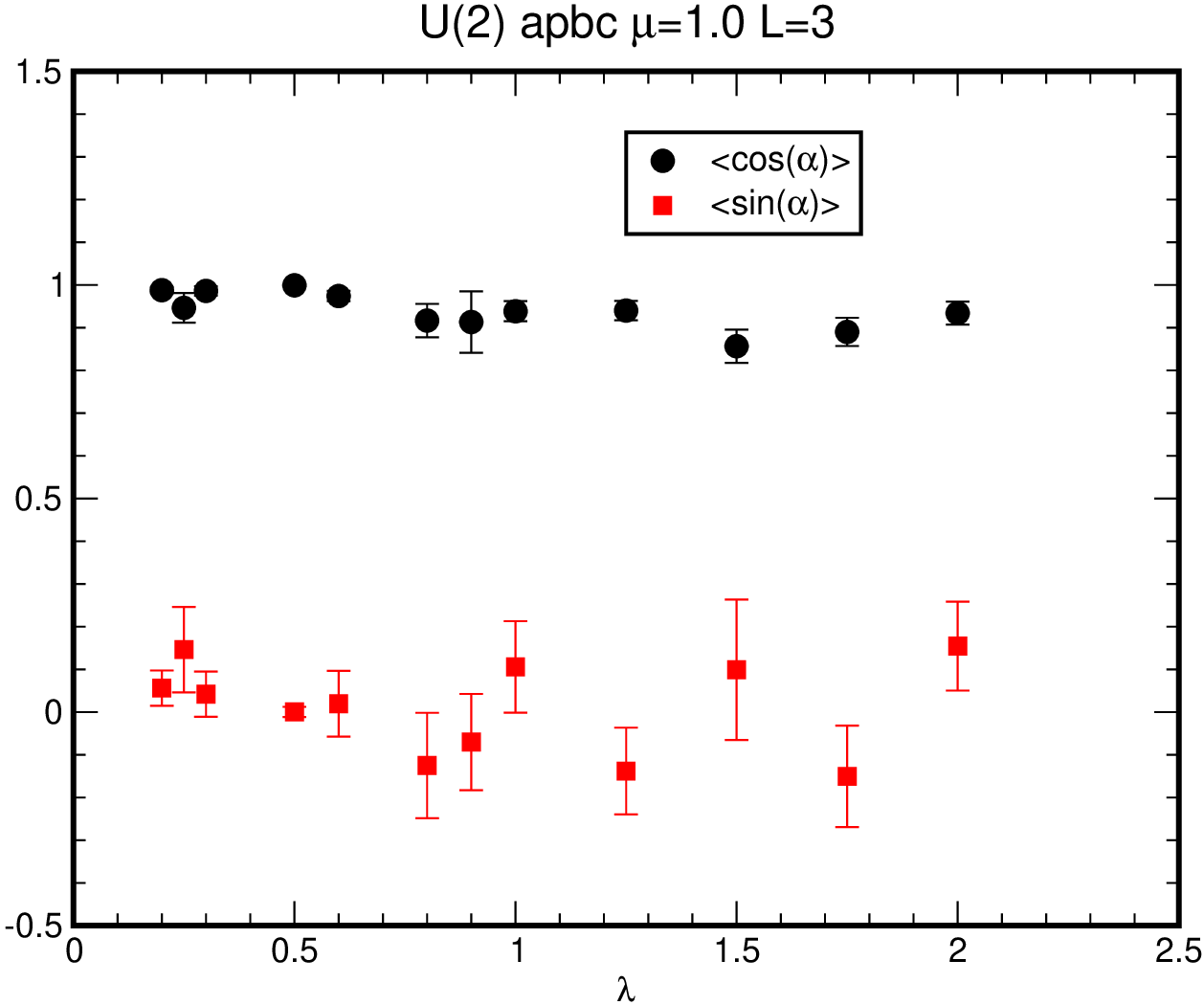}\end{center}
\caption{\label{fig:pfaffian} $ \cos{\alpha}$ and $\sin{\alpha}$ vs $\lambda$ for $\mu_L=1,L=3$.}
\end{figure}

We have also measured the Pfaffian on this small lattice as an explicit check that of possible
sign problems. Fig.~\ref{fig:pfaffian} shows the expectation value of both the cosine and sine of
the Pfaffian phase as a function of $\lambda$. Rather reassuringly we see that the fluctuations
in $\alpha$ are relatively small which provides a concrete numerical justification of the use the
phase quenched approximation in our calculations independent of the measurement of the
bosonic action or analytic arguments based on the topological character of the lattice
partition function.

\subsection{The continuum limit}

Finally, we should address the issue of a continuum limit. If indeed this theory
is conformal at all values of the bare coupling, the beta function vanishes to
all orders in lattice perturbation theory, just as in the continuum\footnote{In \cite{Catterall:2011pd} this
is shown to be true at one loop}. If correct,
this means that the notion of a ``bare'' gauge coupling takes on a new meaning:
a renormalization group flow is not induced by changing the lattice spacing.
Instead, the continuum limit can be reached anywhere on the real positive
bare $g^2$-axis. What about lattice spacing artifacts? The simple way to eliminate
these ultraviolet effects is to go to large distances (volume). In this sense,
detailed simulations of this theory will be highly unusual, much like the
classical solution of differential equations by means of finite differences.
This of course will not mean that the theory is free: there will be anomalous
dimensions and logarithmic behavior beyond classical scaling.

Perhaps it is useful to contrast this situation with ones which are more
familiar to the lattice practitioner. Begin with a pure non-Abelian gauge theory,
defined with an ultraviolet cutoff, the lattice spacing $a$.
It possesses a Gaussian fixed point at $g^2=0$ which is marginally relevant
or unstable under flows towards the infrared. To take the continuum limit,
one must tune the bare coupling to zero. In that limit, correlation lengths
$\xi$ (inverse masses of bound states) become large compared to $a$.
These theories are confining, have a mass gap, and correlation functions always
decay exponentially with distance. One can observe the approach to the
continuum limit in the value of dimensionless ratios
of dimensionful quantities (such as mass ratios).
Any lattice discretization of such a system will possess additional irrelevant
operators. They will affect the spectrum,
and hence the mass ratios. However, these additional irrelevant operators
automatically cease to affect observables as the bare coupling is taken
arbitrarily close to the Gaussian fixed point.
Because these theories have a mass gap, a finite simulation volume (length $L$)
typically affects observables by an amount proportional to $\exp(-mL)$ where $m$
is some characteristic mass.

Next, consider theories of fermions and gauge bosons
``inside the conformal window'', where the gauge coupling
flows to an infrared fixed point under blocking transformations which
remove ultraviolet degrees of freedom. For such theories,
the fermion mass is a relevant perturbation and it must be tuned to zero
by hand in order that the system approach this fixed point. In the massless
limit, and in infinite volume, all correlation functions are power-law
and the interesting physical parameters are the critical exponents.
The distance of the gauge coupling $g$ from its
fixed point value $g_c$ is an irrelevant coupling and the difference
$|g-g_c|$ governs power law corrections to scaling,
whose size is not universal. These effects -- as well as those of all
irrelevant operators -- die away as the correlation length $\xi$ becomes
much greater than the cutoff $a$. Besides the mass, a finite system size
(technically, $1/L$) is also a relevant parameter because it converts
the power-law fall-off of correlation functions into an exponential
fall-off. A combination of simulations done at small but non-zero values
of the relevant parameter (here the fermion mass) and finite simulation volume
(finite size scaling) can, in principle, elucidate the properties of the system.

A theory with a totally vanishing beta function for all bare couplings is one step further. Let
us assume that this is the case for the theory under study here.
This means that when all relevant couplings in the lattice model --
presumably a subset of them are $\mu_L$ and $1/L$ -- are tuned to their
fixed point values, the system will again exhibit algebraic decay of
correlation functions at large distance. This time, the appropriate exponents will be functions
of the bare 't Hooft coupling $\lambda$.

\section{Conclusions}

We have performed numerical simulations of the phase-quenched $\cN=4$
supersymmetric Yang-Mills theory
in four dimensions. In particular, we have
examined standard physical observables such as Wilson loops, Polyakov lines
and the bosonic action. We have found no evidence for
phase transitions as the bare gauge coupling is varied. Furthermore, the effective
string tension is
consistent with zero for all bare couplings at the largest distances probed.
Indeed we see evidence for Coulomb-like behavior in the static quark potential and a gap
opens up in the spectrum of the fermion operator indicating the absence of chiral symmetry
breaking.
Furthermore, the expectation value of the bosonic action appears to be independent
of the gauge coupling as the regulator mass $\mu_L$ is sent to zero and it equals the
value expected on the basis of exact supersymmetry. This gives indirect
evidence that the sign problem is indeed absent in this lattice theory, and
that for all practical purposes the phase of the Pfaffian can be ignored in
actual simulations.

With this first study we have provided ample evidence that it is
feasible to study this supersymmetric lattice gauge theory by numerical means.
The effects of phase quenching, a bosonic mass term to stabilize the flat
directions, and anti-periodic boundary
conditions for the fermions, can all be carefully monitored by means of
supersymmetric Ward Identities that are exact in the lattice-regularized
theory. The apparent existence of a single deconfined phase with vanishing
string tension at all bare couplings indicates that this theory is conformal
at any coupling, as is the continuum theory with all the remaining supercharges
being conserved. Apparently the exact conservation of one supersymmetric
charge is extremely powerful; in particular, it ensures a perfect match
between bosonic and fermionic degrees of freedom in the multiplet.
We hope this work may stimulate renewed numerical efforts
in the same direction: the lattice offers direct access to the computation
of observables in the most interesting supersymmetric gauge theory in
four dimensions. It can probe both weak and strong coupling, and
comparisons can be made to predictions from the continuum based on
gauge-gravity duality.

Our results presented here  are of course only a beginning and should be confirmed by future
studies on bigger systems and at stronger coupling.
The evaluation of non-trivial correlation functions should
be initiated. A study of the broken Ward Identities associated with
supercharges that are not exactly conserved on the lattice should made.
This will give direct evidence for how full supersymmetry is recovered
in the continuum limit. There is obviously much exciting work ahead.

\begin{acknowledgments}
This work was supported by the U.S. Department of Energy grant under
contract nos. DE-FG02-85ER40237 and DE-FG02-04ER41290.
Simulations were performed using USQCD resources at Fermilab and at the
Niels Bohr Institute. Two of us
(SC and PHD) thank the KITP program ``Novel numerical approaches for Strongly Coupled Field Theory and Gravity''
for hospitality during the early stages of this work
and TD thanks the Niels Bohr Institute for its hospitality. One of us (PHD) would like to thank the participants
of the "New Frontiers in Gauge Theory" workshop at
the GGI in Florence for many interesting questions and
discussions.

\end{acknowledgments}

\bibliographystyle{JHEP}

\end{document}